\documentclass[prd,a4paper,showpacs]{revtex4}
\usepackage{epsfig}

\usepackage{amssymb}
\usepackage{amsfonts}
\usepackage{graphicx}
\usepackage{keyval,graphicx}
\usepackage{textcomp,wasysym}
\begin{document}

\bibliographystyle{unsrt}

\title{Further study of the helicity selection rule evading mechanism in $\eta_c$, $\chi_{c0}$ and $h_c$ decaying to baryon anti-baryon pairs}

\author{Xiao-Hai Liu$^1$\footnote{xhliu@ihep.ac.cn}, Qiang Zhao$^{1,2}$\footnote{zhaoq@ihep.ac.cn}}

\affiliation{1) Institute of High Energy Physics, Chinese Academy of
Sciences, Beijing 100049, P.R. China \\
2) Theoretical Physics Center for Science Facilities, CAS, Beijing
100049, China}

\date{\today}

\begin{abstract}
We investigate the long distance contribution via charmed hadron
loops in the processes $\eta_c$, $\chi_{c0}$ and $h_c$ decaying to
baryon anti-baryon pairs, which are supposed to be highly suppressed
by the helicity selection rule as a consequence of the perturbative
QCD framework. With an effective Lagrangian method, our estimation
result indicates that such hadron loops play an important role in
these hadronic decays. It is a further test of the evading mechanism
for the helicity selection rule in charmonium baryon-antibaryon
decays.

\end{abstract}

\maketitle

\section{Introduction}

The complexity of quantum chromodynamics (QCD) remains unsolved in
many questions in the charmonium mass region. This is the regime
that numerous contradictory results between perturbative
calculations and experimental observations were found. In
particular, the pQCD expected helicity selection
rule~\cite{Brodsky:1981kj,Chernyak:1981zz,Chernyak:1983ej} has been
found violated in many exclusive decay processes.  According to this
selection rule, some charmonium decay channels are supposed to be
highly suppressed, such as $J/\psi\to VP$, $\eta_c\to VV$ and
$\chi_{c1}\to VV$, where $V$ and $P$ denote vector and pseudoscalar
meson respectively. However, all these decays are found to be rather
important in experiments \cite{Amsler:2008zzb}. In
Ref.~\cite{Feldmann:2000hs}, Feldmann and Kroll argued that
charmonium mesonic decays could be classified into two catalogues.
One is controlled by pQCD with conserved hadronic helicity while the
other is characterized by the helicity-selection-rule violation. It
was proposed that a soft mechanism through a light-quark Fock
component would account for the violation of the hadronic helicity
conservation in exclusive decays. Other theoretical solutions for
the underlying dynamics have also been explored in the
literature~\cite{Anselmino:1990vs,Feldmann:2000hs,Zhou:2005fc,Zhang:2009kr,Liu:2010chi,Chen:2009ah},
and detailed review of some of the related questions in charmonium
exclusive decays can be found in
Refs.~\cite{Brodsky:1981kj,Chernyak:1981zz,Chernyak:1983ej,Brambilla:2004wf,Voloshin:2007dx,bes-iii}.

In this work we are going to investigate another three processes,
i.e., $\eta_c, \chi_{c0}, h_c\to Y\bar{Y}$, where $Y\bar{Y}$
represent the $J^P=1/2^+$ octet baryon-antibaryon pairs, i.e.
$p\bar{p}$, $\Lambda\bar{\Lambda}$, $\Sigma\bar{\Sigma}$, and
$\Xi\bar{\Xi}$. These channels are also supposed to be highly
suppressed according to the helicity selection rule
\cite{Andrikopoulou:1983sr}. However, it seems that the available
experimental data do not support such expectations at all
\cite{Amsler:2008zzb}. Some attempts have been made in order to
understand this contradiction
\cite{Anselmino:1987du,Anselmino:1991mc,Anselmino:1992jd,Murgia:1996bh,Anselmino:1993yg,Ping:2004sh}.
For instance, in the processes of $\eta_c$ and $\chi_{cJ}(J=0,1,2)$
decaying into $p\bar{p}$, a quark-diquark model for proton is
introduced as a mechanism for evading the helicity selection rule.
However, the obtained branching ratios for $\eta_c$ and $\chi_{cJ}$
can not simultaneously agree with the experimental data
\cite{Anselmino:1987du,Anselmino:1991mc}. Constituent quark mass
corrections were also introduced to account for the processes
$\eta_c, \chi_{c0}, h_c \to p\bar{p}$
\cite{Anselmino:1992jd,Murgia:1996bh}, where it turns out that the
obtained branching ratio for $\eta_c\to p\bar{p}$ is still much
smaller than the experimental data. There are also some other
proposed mechanisms for understanding these processes, such as the
mixing between charmonium state and glueball
\cite{Anselmino:1993yg}, and the quark pair creation model
\cite{Ping:2004sh}. Basically, it is believed that non-perturbative
mechanisms should be relevant in this scenario.

In Refs.~\cite{Zhang:2009kr,Liu:2010chi}, it was proposed that
intermediate meson loop (IML) transitions can serve as a soft
mechanism in charmonium decays. Such a long-distance interaction can
evade the Okubo-Zweig-Iizuka (OZI) rule and result in violation of
the pQCD helicity selection rule. In the framework of effective
Lagrangians for hadron interactions, we can then quantitatively
study the evasion of the helicity selection rule in various
processes.  As a further check of this mechanism, we are going to
investigate the role of the charmed hadron loops in $\eta_c,
\chi_{c0}, h_c\to Y\bar{Y}$ in this work. Additional evidence for
such an underlying mechanism should allow us to gain more insights
into the QCD strong interaction properties in the intermediate
energy region.

The rest of this article is arranged as follows: In Sec. II, we will
depict the effective Lagrangian method, and some relevant formulas
will be given. In Sec. III, we will present the numerical results
and discussions. The conclusions will be summarized in Sec. IV.

\section{Long-distance contribution via charmed hadron loops}

\begin{figure}[htb]
\includegraphics[width=0.5\hsize]{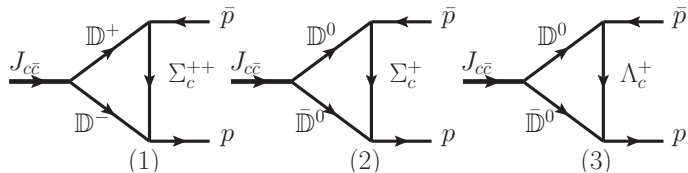}
\caption{Charmed hadron loop diagrams that describe the
long-distance transitions in $J_{c\bar{c}}\to
p\bar{p}$.}\label{fig1}
\end{figure}

\begin{figure}[htb]
\includegraphics[width=0.5\hsize]{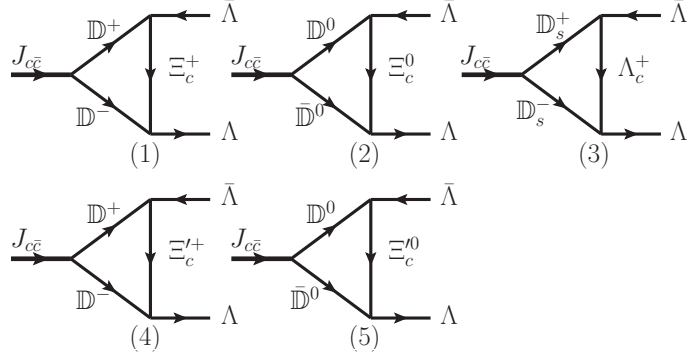}
\caption{Charmed hadron loop diagrams that describe the
long-distance transitions in $J_{c\bar{c}}\to
\Lambda\bar{\Lambda}$.}\label{fig2}
\end{figure}

\begin{figure}[htb]
\includegraphics[width=0.5\hsize]{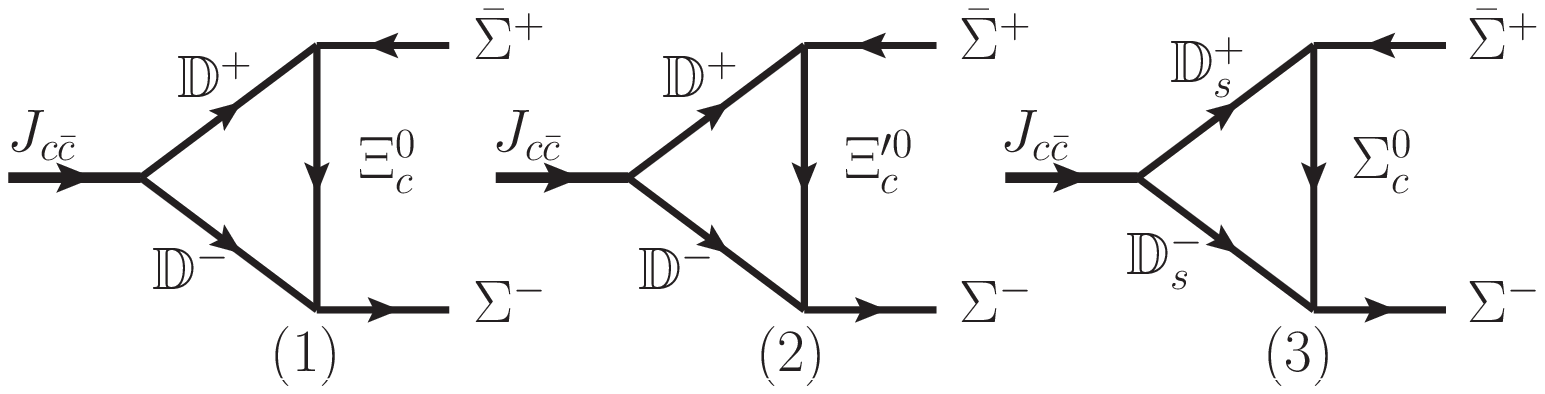}
\caption{Charmed hadron loop diagrams that describe the
long-distance transitions in $J_{c\bar{c}}\to
\Sigma^-\bar{\Sigma}^+$.}\label{fig3}
\end{figure}

\begin{figure}[htb]
\includegraphics[width=0.5\hsize]{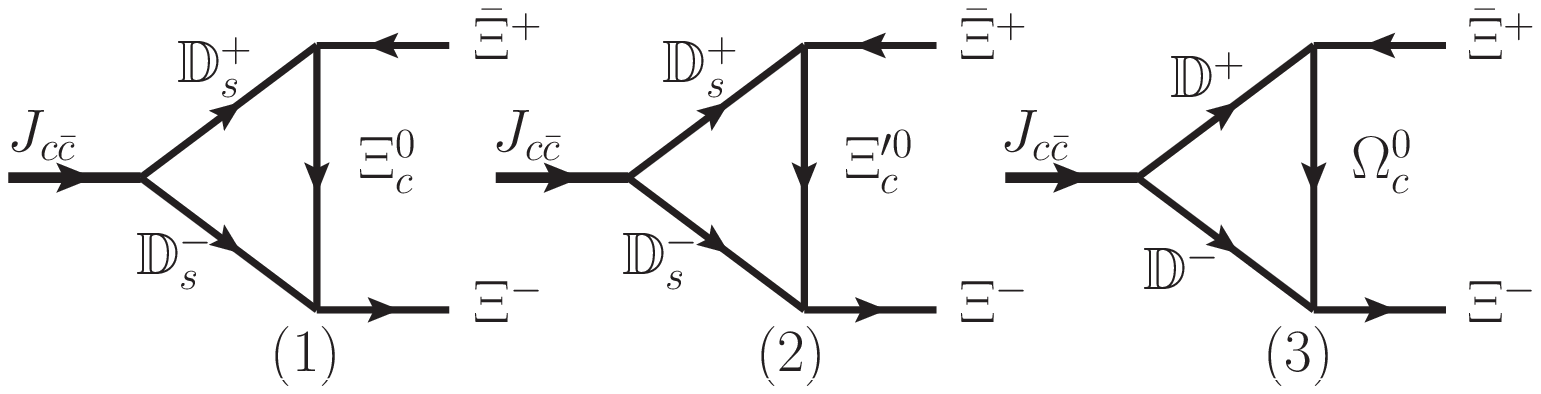}
\caption{Charmed hadron loop diagrams that describe the
long-distance transitions in $J_{c\bar{c}}\to \Xi^-
\bar{\Xi}^+$.}\label{fig4}
\end{figure}

In Figs.~\ref{fig1}, \ref{fig2}, \ref{fig3} and \ref{fig4}, the
intermediate charmed hadron loops which serve as a long-distance
soft mechanism are illustrated by triangle diagrams. Since one
charmed baryon will be present in the loop, these diagrams are
somewhat different from the intermediate meson loops studied in
Refs.~\cite{Zhang:2009kr,Liu:2010chi}. We will consider the
exchanges of the ground state $J^P=1/2^+$ charmed baryons that
belong to SU(4) multiplets~\cite{Amsler:2008zzb,Klempt:2009pi}.
There are several points that should be clarified for
Figs.~\ref{fig1}, \ref{fig2}, \ref{fig3} and \ref{fig4}:

1) $J_{c\bar{c}}$ represents the decaying $c\bar{c}$ meson, i.e.
$\eta_{c}$, $\chi_{c0}$, or $h_c$.

2) $\mathbb{D}_{(s)}$ and $\bar{\mathbb{D}}_{(s)}$ only denote the
flavor contents but do not contain the spin quantum numbers, and
they will depend on the decaying mesons.

3) Different intermediate charmed hadrons will appear in the loops
for $\eta_c, \chi_{c0}, h_c\to Y\bar{Y}$. For the $\eta_c$ decay,
there are three situations for $\mathbb{D}_{(s)}$ and
$\bar{\mathbb{D}}_{(s)}$: (a) $D_{(s)}\bar{D}_{(s)}^*$, (b)
$D_{(s)}^*\bar{D}_{(s)}$, (c) $D_{(s)}^*\bar{D}_{(s)}^*$. For the
$\chi_{c0}$ decay there are two: (a) $D_{(s)}\bar{D}_{(s)}$, (b)
$D_{(s)}^*\bar{D}_{(s)}^*$. For the $h_c$ decay there are three: (a)
$D_{(s)}\bar{D}_{(s)}^*$, (b) $D_{(s)}^*\bar{D}_{(s)}$, (c)
$D_{(s)}^*\bar{D}_{(s)}^*$. This is due to the adopted effective
Lagrangians based on heavy quark symmetry
\cite{Colangelo:2003sa,Casalbuoni:1996pg}.

We list the relevant effective Lagrangians as follows:
\begin{eqnarray}
\mathcal{L}_1 &=& i g_1 Tr[P_{c\bar{c}}^\mu \bar{H}_{2i}\gamma_\mu
\bar{H}_{1i}] + h.c., \\
\mathcal{L}_2 &=& i g_2 Tr[R_{c\bar{c}} \bar{H}_{2i}\gamma^\mu
{\stackrel{\leftrightarrow}{\partial}}_\mu \bar{H}_{1i}] + h.c.,
\end{eqnarray}
where the spin multiplets for these four $P$-wave and two $S$-wave
charmonium states are expressed as
\begin{eqnarray}
P_{c\bar{c}}^\mu &=& \left( \frac{1+ \rlap{/}{v} }{2} \right)
\left(\chi_{c2}^{\mu\alpha}\gamma_{\alpha} +\frac{1}{\sqrt{2}}
\epsilon^{\mu\nu\alpha\beta}v_{\alpha}\gamma_{\beta}\chi_{c1\nu}
+\frac{1}{\sqrt{3}}(\gamma^\mu -v^\mu)\chi_{c0} +h_c^\mu \gamma_{5}
\right) \left( \frac{1- \rlap{/}{v} }{2} \right), \\ \label{pwave}
R_{c\bar{c}}&=& \left( \frac{1+ \rlap{/}{v} }{2} \right) (\psi^\mu
\gamma_\mu-\eta_c \gamma_5) \left( \frac{1- \rlap{/}{v} }{2}
\right).
\end{eqnarray}
The charmed and anti-charmed meson triplet read
\begin{eqnarray}
H_{1i}&=&\left( \frac{1+ \rlap{/}{v} }{2} \right)
[\mathcal{D}_i^{*\mu}
\gamma_\mu -\mathcal{D}_i\gamma_5], \\
H_{2i}&=& [\bar{\mathcal{D}}_i^{*\mu} \gamma_\mu
-\bar{\mathcal{D}}_i\gamma_5]\left( \frac{1- \rlap{/}{v} }{2}
\right),
\end{eqnarray}
where $\mathcal{D}$ and $\mathcal{D}^*$ denote the pseudoscalar and
vector charmed meson fields respectively, i.e.
$\mathcal{D}^{(*)}=(D^{0(*)},D^{+(*)},D_s^{+(*)})$. For the
meson-baryon interaction Lagrangians, we follow the forms that were
adopted in
Refs.~\cite{Lin:1999ve,Haglin:1999xs,Rekalo:2001pp,Rekalo:2000hh}:
\begin{eqnarray}
\mathcal{L}_{Y_c\mathcal{D}Y} &=& i g_{Y_c\mathcal{D}Y} \bar{Y}_c
\gamma_5 Y \mathcal{D}+h.c., \\
\mathcal{L}_{Y_c\mathcal{D}^*Y} &=& g_{Y_c\mathcal{D}^*Y}
\bar{Y}_c\left( \gamma^\mu \mathcal{D}^*_\mu +
\frac{\kappa_{Y_c\mathcal{D}^*Y}}{2m_N} \sigma^{\mu\nu}\partial_\mu
\mathcal{D}^*_\nu \right) Y +h.c.,
\end{eqnarray}
where $Y_c$, $\mathcal{D}^{(*)}$, and $Y$ denote the charmed baryon,
charmed meson, and the corresponding nucleon or hyperon,
respectively. The relevant coupling constants will be discussed
later.

With the above effective Lagrangians, we can now calculate the
transition amplitudes illustrated in Figs.~\ref{fig1}, \ref{fig2},
\ref{fig3} and \ref{fig4}. For these diagrams, we take the
convention of the momenta as $J_{c\bar{c}}(p)\to
\mathbb{D}(q_1)\bar{\mathbb{D}}(q_2)[Y_c(q)]\to \bar{Y}(p_1)
Y(p_2)$.

\subsection{$\eta_c\to Y\bar{Y}$}
We first consider $\eta_{c}\to Y\bar{Y}$, where three different
amplitudes corresponding to those three intermediate states will
contribute, i.e. $\mathcal{M}_a$, $\mathcal{M}_b$, and
$\mathcal{M}_c$. These amplitudes can be written down explicitly as
follows:
\begin{eqnarray}
\mathcal{M}_a &=& 2 g_{\eta_c \mathcal{D}\mathcal{D}^*} g_{Y_c
\mathcal{D} Y} g_{Y_c\mathcal{D}^*Y}
 \int\frac{d^4q}{(2\pi)^4}(q_{2\lambda}-q_{1\lambda})
 \left(-g^{\lambda\mu}+\frac{q_2^\lambda
 q_2^\mu}{m_{\mathcal{D}^*}^2} \right) \nonumber \\
& \times & \bar{u}(p_2) \left(\gamma_\mu+i
\frac{\kappa_{Y_c\mathcal{D}^*Y}}{2m_N} \sigma_{\mu\nu} q_{2}^\nu
\right) (\rlap{/}{q} + m_{Y_c}) \gamma_5 v(p_1) \nonumber \\
&\times&
\frac{1}{q^2-m_{Y_c}^2}\frac{1}{q_1^2-m_{\mathcal{D}}^2}\frac{1}{q_2^2-m_{\mathcal{D}^*}^2}
\mathcal{F}(q^2), \\
\mathcal{M}_b &=& 2 g_{\eta_c \mathcal{D}\mathcal{D}^*} g_{Y_c
\mathcal{D} Y} g_{Y_c\mathcal{D}^*Y}
 \int\frac{d^4q}{(2\pi)^4}(q_{2\lambda}-q_{1\lambda})
 \left(-g^{\lambda\mu}+\frac{q_1^\lambda
 q_1^\mu}{m_{\mathcal{D}^*}^2} \right) \nonumber \\
& \times & \bar{u}(p_2)\gamma_5  (\rlap{/}{q} + m_{Y_c})
\left(\gamma_\mu+i \frac{\kappa_{Y_c\mathcal{D}^*Y}}{2m_N}
\sigma_{\mu\nu} q_{1}^\nu
\right)  v(p_1) \nonumber \\
&\times&
\frac{1}{q^2-m_{Y_c}^2}\frac{1}{q_1^2-m_{\mathcal{D}^*}^2}\frac{1}{q_2^2-m_{\mathcal{D}}^2}
\mathcal{F}(q^2), \\
\mathcal{M}_c &=& -2i g_{\eta_c
\mathcal{D}^*\mathcal{D}^*}g_{Y_c\mathcal{D}^*Y}^2
 \int\frac{d^4q}{(2\pi)^4} \epsilon^{\mu\nu\lambda\tau} p_\nu
 (q_{2\mu}-q_{1\mu}) \nonumber \\
& \times & \bar{u}(p_2) \left(\gamma_\tau+i
\frac{\kappa_{Y_c\mathcal{D}^*Y}}{2m_N} \sigma_{\tau\xi} q_{2}^\xi
\right) (\rlap{/}{q} + m_{Y_c}) \left(\gamma_\lambda+i
\frac{\kappa_{Y_c\mathcal{D}^*Y}}{2m_N} \sigma_{\lambda\sigma}
q_{1}^\sigma
\right) v(p_1) \nonumber \\
&\times&
\frac{1}{q^2-m_{Y_c}^2}\frac{1}{q_1^2-m_{\mathcal{D}^*}^2}\frac{1}{q_2^2-m_{\mathcal{D}^*}^2}
\mathcal{F}(q^2).
\end{eqnarray}

Since the masses of $\eta_c, \chi_{c0}$, and $h_c$ are below the
threshold of $D^{(*)}\bar{D}^{(*)}$, intermediate mesons $D^{(*)}$
and $\bar{D}^{(*)}$ can not be on-shell simultaneously. We thus
phenomenologically introduce a form factor $\mathcal{F}(q^2)$ as has
been done in Refs. \cite{Liu:2010chi,Colangelo:2003sa,Cheng:2004ru}
to take into account the off-shell effects,
\begin{eqnarray}
\mathcal{F}(q^2)=\prod\limits_i
\left(\frac{m_i^2-\Lambda_i^2}{q_i^2-\Lambda_i^2} \right),
\end{eqnarray}
where $q_i=q,\ q_1,\ q_2$. The cut-off energy is chosen as
$\Lambda_i=m_i +\alpha \Lambda_{QCD}$, $\Lambda_{QCD}=0.22\
\mbox{GeV}$ and $m_i$ is the mass of the corresponding exchanged
particle. The form factor is also necessary for killing the
divergence of the loop integrals, although it will give rise to
model-dependent aspects of the calculations. Qualitatively, this
type of form factor will converge the integrals faster than a dipole
form factor and can be well combined with a vertex coupling constant
as a phenomenological account of the non-local coupling form factor.
Some further discussions will be given later in the numerical
analysis.

In this work, we do not include contributions from exchanging other
charmed hadrons with higher spin or orbital excitations. These
resonances have relatively larger masses and their couplings so far
are unknown. It should be reasonable to only consider the lowest
partial wave states based on the argument of locally broken down
quark-hadron duality~\cite{Li:2007xr}.

\subsection{$\chi_{c0}\to Y\bar{Y}$}

Two charmed hadron loops will contribute to $\chi_{c0}\to Y\bar{Y}$,
and the corresponding amplitudes are
\begin{eqnarray}
\mathcal{M}_a &=& -ig_{\chi_{c0} \mathcal{D}\mathcal{D}} g_{Y_c
\mathcal{D} Y}^2
 \int\frac{d^4q}{(2\pi)^4}
 \bar{u}(p_2)\gamma_5  (\rlap{/}{q} + m_{Y_c}) \gamma_5 v(p_1) \nonumber \\
&\times&
\frac{1}{q^2-m_{Y_c}^2}\frac{1}{q_1^2-m_{\mathcal{D}}^2}\frac{1}{q_2^2-m_{\mathcal{D}}^2}
\mathcal{F}(q^2), \\
\mathcal{M}_b &=& i g_{\chi_{c0}
\mathcal{D}^*\mathcal{D}^*}g_{Y_c\mathcal{D}^*Y}^2
 \int\frac{d^4q}{(2\pi)^4} g_{\mu\nu}
 \left(-g^{\mu\rho}+\frac{q_1^\mu
 q_1^\rho}{m_{\mathcal{D}^*}^2} \right)
 \left(-g^{\nu\alpha}+\frac{q_2^\nu
 q_2^\alpha}{m_{\mathcal{D}^*}^2} \right) \nonumber \\
& \times & \bar{u}(p_2) \left(\gamma_\alpha+i
\frac{\kappa_{Y_c\mathcal{D}^*Y}}{2m_N} \sigma_{\alpha\beta}
q_{2}^\beta \right) (\rlap{/}{q} + m_{Y_c}) \left(\gamma_\rho+i
\frac{\kappa_{Y_c\mathcal{D}^*Y}}{2m_N} \sigma_{\rho\tau} q_{1}^\tau
\right) v(p_1) \nonumber \\
&\times&
\frac{1}{q^2-m_{Y_c}^2}\frac{1}{q_1^2-m_{\mathcal{D}^*}^2}\frac{1}{q_2^2-m_{\mathcal{D}^*}^2}
\mathcal{F}(q^2),
\end{eqnarray}
where the form factor has the same form as that in $\eta_c\to
Y\bar{Y}$.

\subsection{$h_c\to Y\bar{Y}$}

Similarly, the amplitudes for $h_c\to Y\bar{Y}$ from those three
contributing loops can be written down as follows,
\begin{eqnarray}
\mathcal{M}_a &=& g_{h_c \mathcal{D}\mathcal{D}^*} g_{Y_c
\mathcal{D} Y} g_{Y_c\mathcal{D}^*Y} \epsilon_\eta(p)
 \int\frac{d^4q}{(2\pi)^4}
 \left(-g^{\lambda\mu}+\frac{q_2^\lambda
 q_2^\mu}{m_{\mathcal{D}^*}^2} \right) \nonumber \\
& \times & \bar{u}(p_2) \left(\gamma_\mu+i
\frac{\kappa_{Y_c\mathcal{D}^*Y}}{2m_N} \sigma_{\mu\nu} q_{2}^\nu
\right) (\rlap{/}{q} + m_{Y_c}) \gamma_5 v(p_1) \nonumber \\
&\times&
\frac{1}{q^2-m_{Y_c}^2}\frac{1}{q_1^2-m_{\mathcal{D}}^2}\frac{1}{q_2^2-m_{\mathcal{D}^*}^2}
\mathcal{F}(q^2), \\
\mathcal{M}_b &=& g_{h_c \mathcal{D}\mathcal{D}^*} g_{Y_c
\mathcal{D} Y} g_{Y_c\mathcal{D}^*Y} \epsilon_\eta(p)
 \int\frac{d^4q}{(2\pi)^4}
 \left(-g^{\lambda\mu}+\frac{q_1^\lambda
 q_1^\mu}{m_{\mathcal{D}^*}^2} \right) \nonumber \\
& \times & \bar{u}(p_2)\gamma_5  (\rlap{/}{q} + m_{Y_c})
\left(\gamma_\mu+i \frac{\kappa_{Y_c\mathcal{D}^*Y}}{2m_N}
\sigma_{\mu\nu} q_{1}^\nu
\right)  v(p_1) \nonumber \\
&\times&
\frac{1}{q^2-m_{Y_c}^2}\frac{1}{q_1^2-m_{\mathcal{D}^*}^2}\frac{1}{q_2^2-m_{\mathcal{D}}^2}
\mathcal{F}(q^2), \\
\mathcal{M}_c &=&  g_{h_c
\mathcal{D}^*\mathcal{D}^*}g_{Y_c\mathcal{D}^*Y}^2 \epsilon^\eta(p)
 \int\frac{d^4q}{(2\pi)^4} \epsilon_{\rho\eta\alpha\beta} p^\rho
  \left(-g^{\alpha\lambda}+\frac{q_1^\alpha
 q_1^\lambda}{m_{\mathcal{D}^*}^2} \right)
 \left(-g^{\beta\tau}+\frac{q_2^\beta
 q_2^\tau}{m_{\mathcal{D}^*}^2} \right) \nonumber \\
& \times & \bar{u}(p_2) \left(\gamma_\tau+i
\frac{\kappa_{Y_c\mathcal{D}^*Y}}{2m_N} \sigma_{\tau\xi} q_{2}^\xi
\right) (\rlap{/}{q} + m_{Y_c}) \left(\gamma_\lambda+i
\frac{\kappa_{Y_c\mathcal{D}^*Y}}{2m_N} \sigma_{\lambda\sigma}
q_{1}^\sigma
\right) v(p_1) \nonumber \\
&\times&
\frac{1}{q^2-m_{Y_c}^2}\frac{1}{q_1^2-m_{\mathcal{D}^*}^2}\frac{1}{q_2^2-m_{\mathcal{D}^*}^2}
\mathcal{F}(q^2),
\end{eqnarray}
where $\epsilon_\eta(p)$ is the polarization vector for $h_c$, and
again the form factor has the same form as that in $\eta_c\to
Y\bar{Y}$.

\section{Numerical Results and Discussion}

Proceeding to the numerical results, we first discuss the
determination of coupling constants. For the couplings of the
charmonium states to charmed mesons, the expansion of the effective
Lagrangians $\mathcal{L}_1$ and $\mathcal{L}_2$ gives the following
relations in the heavy quark limit:
\begin{eqnarray}
g_{\eta_c \mathcal{D}\mathcal{D}^*} &=& 2 g_2
\sqrt{m_{\eta_c}m_{\mathcal{D}}m_{\mathcal{D}^*}},\ g_{\eta_c
\mathcal{D}^*\mathcal{D}^*} =2 g_2
\frac{m_{\mathcal{D}^*}}{\sqrt{m_{\eta_c}}}, \nonumber \\
g_{\chi_{c0}\mathcal{D}\mathcal{D}} &=& -2\sqrt{3} g_1
m_{\mathcal{D}}\sqrt{m_{\chi_{c0}}},\
g_{\chi_{c0}\mathcal{D}^*\mathcal{D}^*}=-\frac{2}{\sqrt{3}} g_1
m_{\mathcal{D}^*}\sqrt{m_{\chi_{c0}}}, \nonumber \\
g_{h_c \mathcal{D}\mathcal{D}^*} &=& -2 g_1
\sqrt{m_{h_c}m_{\mathcal{D}} m_{\mathcal{D}^*}},\
g_{h_c\mathcal{D}^*\mathcal{D}^*} = 2 g_1
\frac{m_{\mathcal{D}^*}}{\sqrt{m_{h_c}}}, \nonumber \\
&& g_1 = -\sqrt{\frac{m_{\chi_{c0}}}{3}} \frac{1}{f_{\chi_{c0}} },\
g_2 =\frac{\sqrt{m_\psi}}{2m_{\mathcal{D}} f_\psi},
\end{eqnarray}
where $f_{\chi_{c0}}$ and $f_{\psi}$ are the decay constants of
$\chi_{c0}$ and $J/\psi$, respectively. $f_\psi\simeq 0.41\
\mbox{GeV}$, which is determined by the experimental data
\cite{Amsler:2008zzb}. And $f_{\chi_{c0}}$ can be approximately
determined by the QCD sum rule approach, i.e. $f_{\chi_{c0}} \simeq
0.51\ \mbox{GeV}$ \cite{Colangelo:2002mj}.

There is no much information on the couplings of a charmed baryon to
a charmed meson and light baryon. If considering SU(4) flavor
symmetry, it will relate different couplings with each other. For
the considered $J^P=1/2^+$ charmed baryons, we would expect the
following relations \cite{Bose:1976ex}:
\begin{eqnarray}\label{su4-relation}
&& [\bar{\Sigma}_c^{--} D^{+} p]= [\bar{\Sigma}_c^{--} D^{+}_s
\Sigma^+] = -\sqrt{2} [{\bar{\Xi}}^{\prime0}_c D^{+} \Sigma^-]
=-\sqrt{2}
[{\bar{\Xi}}^{\prime-}_c D^{+}_s \Xi^0] \nonumber \\
&& = -\frac{2}{\sqrt{3}} [{\bar{\Xi}}^{\prime-}_c D^{+} \Lambda]
=-[\bar{\Omega}_c^0 D^+ \Xi^-] =-\sqrt{2}[\bar{p} K^+ \Sigma^0],
\nonumber \\
&&  [\bar{\Lambda}_c^{-} D^{+} n]= [\bar{\Xi}_c^0 D^{+} \Sigma^-] =
-[\bar{\Xi}_c^- D^{+}_s \Xi^0] =-\sqrt{6} [\bar{\Xi}_c^- D^{+}
\Lambda] \nonumber \\
&& = \sqrt{\frac{3}{2}} [\bar{\Lambda}_c^{-} D^{+}_s \Lambda] =
-[\bar{p} K^+ \Lambda],
\end{eqnarray}
where the square bracket ``$[\cdots]$" denotes the coupling constant
for the corresponding vertex, and the pseudoscalar meson in the
bracket can be replaced by the corresponding vector meson. These
formulas relate the charmed-baryon-meson couplings with the
strange-baryon-meson coupling. For instance, the commonly adopted
value for $\Lambda K N$ coupling is $g_{\Lambda KN}=-13.2$
\cite{Rijken:1998yy,Stoks:1999bz,Oh:2006hm}, from which we obtain
$g_{\Lambda_c D N}=-g_{\Lambda KN}=13.2$. Such a relation may
contain rather large uncertainties as we can see that the QCD sum
rule approach suggests a smaller value $|g_{\Lambda_c D N}|=6.7\pm
2.1$ \cite{Navarra:1998vi}. Experimental data from the charmed meson
photoproduction and charmonium absorption by nucleons may offer some
constraints on these couplings. Unfortunately, the extractions of
the couplings still depend on the adopted theoretical models
\cite{Rekalo:2000hh,Haglin:1999xs,Lin:1999ve}. As a tentative
solution for this, we empirically retain the SU(4) symmetry
relations in the numerical calculation. The relevant
strange-baryon-meson couplings are taken from the Nigmegen potential
model as follows \cite{Rijken:1998yy,Stoks:1999bz}:
\begin{eqnarray}
&&g_{\Lambda K^* N}=-4.26,\ \kappa_{\Lambda K^* N}=2.16, \nonumber \\
&&g_{\Sigma  K^* N}=-2.5,\ \kappa_{\Sigma  K^* N}=-0.22,
\end{eqnarray}
and
\begin{eqnarray}
g_{\Lambda K N}=-13.2,\ g_{\Sigma K N}=3.9.
\end{eqnarray}

The form factor parameter $\alpha$ generally cannot be determined
from the first principle. It will depend on a particular process and
its value is order of unity. There might exist some differences
between the values of the form factor parameter for $\eta_c\to
p\bar{p}$ and $\chi_{c0}\to p\bar{p}$ since $\eta_c$ and $\chi_{c0}$
belong to different spin multiplets. It is natural to anticipate
that the counter term structures would be different for loops
involving different spin multiplets. Similar feature was also found
in Ref.~\cite{Cheng:2004ru} and other studies~\cite{Liu:2010chi}. As
emphasized in the literature, in order to reduce the uncertainties
arising from the form factor, one first has to rely on the
experimental data to determine the form factor parameter, and then
apply it to other flavor-symmetry-related processes to make
predictions which can be examined by further experimental data. This
empirical treatment will also be useful for a further control of the
uncertainties arising from the SU(4) flavor symmetry breaking.
Namely, with the charmed-baryon-meson couplings fixed by the SU(4)
relations, we allow the experimental data for $\eta_c\to p\bar{p}$
to determine the range of $\alpha$ as a compensation of the
uncertainties from the couplings. This can be regarded reasonable
since a physical coupling form factor will generally be correlated
with these two aspects. Reliability of this treatment can be tested
by the prediction for the $\eta_c\to \Lambda\bar{\Lambda}$ branching
ratio, which turns out to be consistent with the experimental
data~\cite{Amsler:2008zzb}.

At present, the branching ratios for $\eta_c\to p\bar{p}$ and
$\chi_{c0}\to p\bar{p}$ have been measured by experiment
\cite{Amsler:2008zzb}. Signals of $h_c$ were also found in
$p\bar{p}$ annihilations \cite{Armstrong:1992ae,Andreotti:2005vu},
which is a hint that $h_c\to p\bar{p}$ could be an important channel
in the $h_c$ decays. Thus, we will use the measured branching ratios
$\mbox{BR}(\eta_c\to p\bar{p})$ and $\mbox{BR}(\chi_{c0}\to
p\bar{p})$ to extract the form factor parameters for the $S$-wave
state $\eta_c$ and $P$-wave state $\chi_{c0}$, respectively.
Considering that $h_c$ belongs to the same spin multiplet as
$\chi_{c0}$, we conjecture that they may share the same intrinsic
dynamics in their decays into $p\bar{p}$. We will then adopt the
same form factor parameter for $h_c$ as that extracted from
$\chi_{c0}\to p\bar{p}$.

We use the software package LoopTools to calculate the loop
integrals \cite{Hahn:1998yk}. The results are displayed in
Tables~\ref{table1}, \ref{table2}, and \ref{table3}. The
experimental data for $\eta_c\to p\bar{p}$ and $\chi_{c0}\to
p\bar{p}$ are adopted for the determination of the form factor
parameter $\alpha$ with the range of the uncertainties. In
Table~\ref{table1}, the predicted branching ratio,
$\mbox{BR}(\eta_c\to  \Lambda\bar{\Lambda})=(6.3\sim 12.5)\times
10^{-4}$, is consistent with the data $\mbox{BR}^{exp}(\eta_c\to
\Lambda\bar{\Lambda})=(10.4\pm 3.1)\times
10^{-4}$~\cite{Amsler:2008zzb}, which is a sign for the parameter
under control. The experimental data for $\eta_c\to
\Sigma\bar{\Sigma}$ and $\Xi\bar{\Xi}$ are unavailable. Our
calculations suggest that these two branching ratios are compatible
with that for $\eta_c\to  \Lambda\bar{\Lambda}$. This expectation
can be examined by the BESIII experiment.

The experimental data for $\chi_{c0}\to p\bar{p}$ and
$\Lambda\bar{\Lambda}$ will allow a further check of the model and
its parameter space in the spin-1 multiplets. In Table~\ref{table2},
the calculation results are given by the form factor parameter
within a range, i.e. $\alpha=0.23\sim 0.24$, which corresponds to
the experimental uncertainties of $\chi_{c0}\to
p\bar{p}$~\cite{Amsler:2008zzb}. This value range is different from
that for the $\eta_c$ decays with $\alpha=0.47\sim 0.53$. This can
be understood as a consequence that $\chi_{c0}$ and $\eta_c$ belong
to different spin multiplets. We shall show later the sensitivities
of the calculation results to the form factor parameter later. For
the $\chi_{c0}$ decays, the data from CLEO suggest relatively larger
branching ratios for $\chi_{c0}\to \Lambda\bar{\Lambda}$ compared
with that for $\chi_{c0}\to p\bar{p}$. Also, branching ratios for
$\chi_{c0}\to \Sigma\bar{\Sigma}$ and $\Xi\bar{\Xi}$ are sizeable.
In contrast, with the same form factor parameter $\alpha$, we find
relatively smaller branching ratios for $\chi_{c0}\to
\Lambda\bar{\Lambda}$, $\Sigma\bar{\Sigma}$ and $\Xi\bar{\Xi}$. It
could be a sign that the SU(4) flavor symmetry is badly broken.
Namely, Eq.~(\ref{su4-relation}) may be too rough, and can only
provide an estimate of magnitude orders for $\chi_{c0}\to
\Sigma\bar{\Sigma}$ and $\Xi\bar{\Xi}$. In this sense, the
calculation results for $\chi_{c0}\to \Lambda\bar{\Lambda}$,
$\Sigma\bar{\Sigma}$ and $\Xi\bar{\Xi}$, though turn out to be
smaller than the experimental data, can be regarded as reasonable.

Although the present model uncertainties do not allow us to conclude
the magnitudes of the hadron loop contributions, the pattern
predicted by this mechanism still suggests an important role played
by the hadron loops in the explanation of the
helicity-selection-rule violations in these exclusive decays.
Meanwhile, we stress that it would be essential to have improved
experimental measurements in order to gain better insights into the
transition mechanism.

\begin{center}
\begin{table}
\begin{tabular}{|c|c|c|c|c|}
 \hline
  \hline
  BR$(\mbox{in units of}\ 10^{-4})$  & $p\bar{p}$ & $\Lambda\bar{\Lambda}$ & $\Sigma^-\bar{\Sigma}^+$ &$\Xi^-\bar{\Xi}^+$ \\
  \hline
  Hadron loop & $9.0\sim 17.0$ & $6.3\sim 12.5$ & $5.05\sim 10.0$ & $4.82\sim 9.56$ \\
  \hline
  Exp. & $13\pm 4$ & $10.4\pm 3.1$ & - & - \\
  \hline
\end{tabular}
\caption{Branching ratios for $\eta_c\to Y\bar{Y}$ predicted by the
intermediate charmed hadron loop transitions in the range
$\alpha=0.47\sim 0.53$ which corresponds to the measured lower and
upper bound of $\mbox{BR}(\eta_c\to p\bar{p})$ . The available
experimental data are taken from Ref.~\cite{Amsler:2008zzb}, and the
dashes mean that the data are unavailable. } \label{table1}
\end{table}
\end{center}

\begin{center}
\begin{table}
\begin{tabular}{|c|c|c|c|c|}
 \hline
  \hline
  BR$(\mbox{in units of}\ 10^{-4})$  & $p\bar{p}$ & $\Lambda\bar{\Lambda}$ & $\Sigma^-\bar{\Sigma}^+$ &$\Xi^-\bar{\Xi}^+$ \\
  \hline
  Hadron loop & $1.96\sim 2.34$ & $1.19\sim 1.51$ & $0.55\sim 0.69$ & $0.52\sim 0.66$ \\
  \hline
  Exp. \cite{Amsler:2008zzb} & $2.15\pm 0.19$ & $4.4\pm 1.5$ & - & $<10.3$ \\
  \hline
  Exp. \cite{Naik:2008dk} & $2.25\pm 0.27$ & $4.7\pm 1.6$ & $3.25\pm 1.14$ & $5.14\pm 1.25$ \\
  \hline
\end{tabular}
\caption{Branching ratios for $\chi_{c0}\to Y\bar{Y}$ predicted by
the intermediate charmed hadron loop transitions in the range
$\alpha=0.23\sim 0.24$ which corresponds to the measured lower and
upper bound of $\mbox{BR}(\chi_{c0}\to p\bar{p})$
\cite{Amsler:2008zzb}.} \label{table2}
\end{table}
\end{center}

\begin{center}
\begin{table}
\begin{tabular}{|c|c|c|c|c|}
 \hline
  \hline
  BR$(\mbox{in units of}\ 10^{-4})$  & $p\bar{p}$ & $\Lambda\bar{\Lambda}$ & $\Sigma^-\bar{\Sigma}^+$ &$\Xi^-\bar{\Xi}^+$ \\
  \hline
  Hadron loop & $15.2\sim 19.3$ & $5.88\sim 7.47$ & $4.56\sim 5.80$ & $5.57\sim 7.08$ \\
  \hline
  Exp. & - & - & - & - \\
  \hline
\end{tabular}
\caption{Branching ratios for $h_c\to Y\bar{Y}$ predicted by the
intermediate charmed hadron loop transitions.  The $\alpha$ range
for $h_c\to Y\bar{Y}$ is taken the same as that for $\chi_{c0}\to
Y\bar{Y}$, i.e. $\alpha=0.23\sim 0.24$. The dashes mean that the
data are unavailable. We take the width of $h_c$ as
$\Gamma(h_c)=0.73 \ \mbox{MeV}$, which is the central value measured
by BESIII recently \cite{Collaboration:2010rc}. } \label{table3}
\end{table}
\end{center}

Experimental data for $h_c\to Y\bar{Y}$ so far are unavailable.
Thus, our predictions are based on the assumption that $h_c$ shares
the same form factor parameter as $\chi_{c0}$ since they belong to
the same spin multiplet. In Table~\ref{table3} we list the branching
ratios given by $\alpha=0.23\sim 0.24$, which is the same as adopted
in $\chi_{c0}\to p\bar{p}$. In comparison with the theoretical
calculations of
Refs.~\cite{Murgia:1996bh,Barnes:2004uc,Kuang:1988bz}, our
prediction of the branching ratio of $h_c\to p\bar{p}$ seems to be
larger. In fact, the theoretical predictions in the literature also
appear to be quite different from each other. In particular, some of
those results strongly depend on the evaluation of $\mbox{BR}(h_c\to
J/\psi \pi^0)$ in the combined cross sections for $p\bar{p}\to
h_c\to J/\psi\pi^0$ from the E760 data
\cite{Murgia:1996bh,Armstrong:1992ae,Barnes:2004uc}. Another reason
for the discrepancies among these theoretical estimates may be due
to different intrinsic mechanisms adopted for the explanation of the
helicity-selection-rule violation. We expect that the future precise
measurement of $h_c\to Y\bar{Y}$ will help disentangle the
underlying mechanisms.

\begin{figure}[htb]
\includegraphics[width=0.5\hsize]{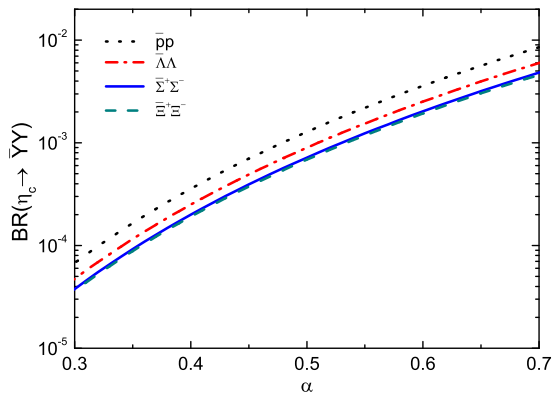}
\caption{ $\alpha$-dependence of the calculated branching ratios for
$\eta_c\to Y\bar{Y}$. }\label{fig5}
\end{figure}

\begin{figure}[htb]
\includegraphics[width=0.5\hsize]{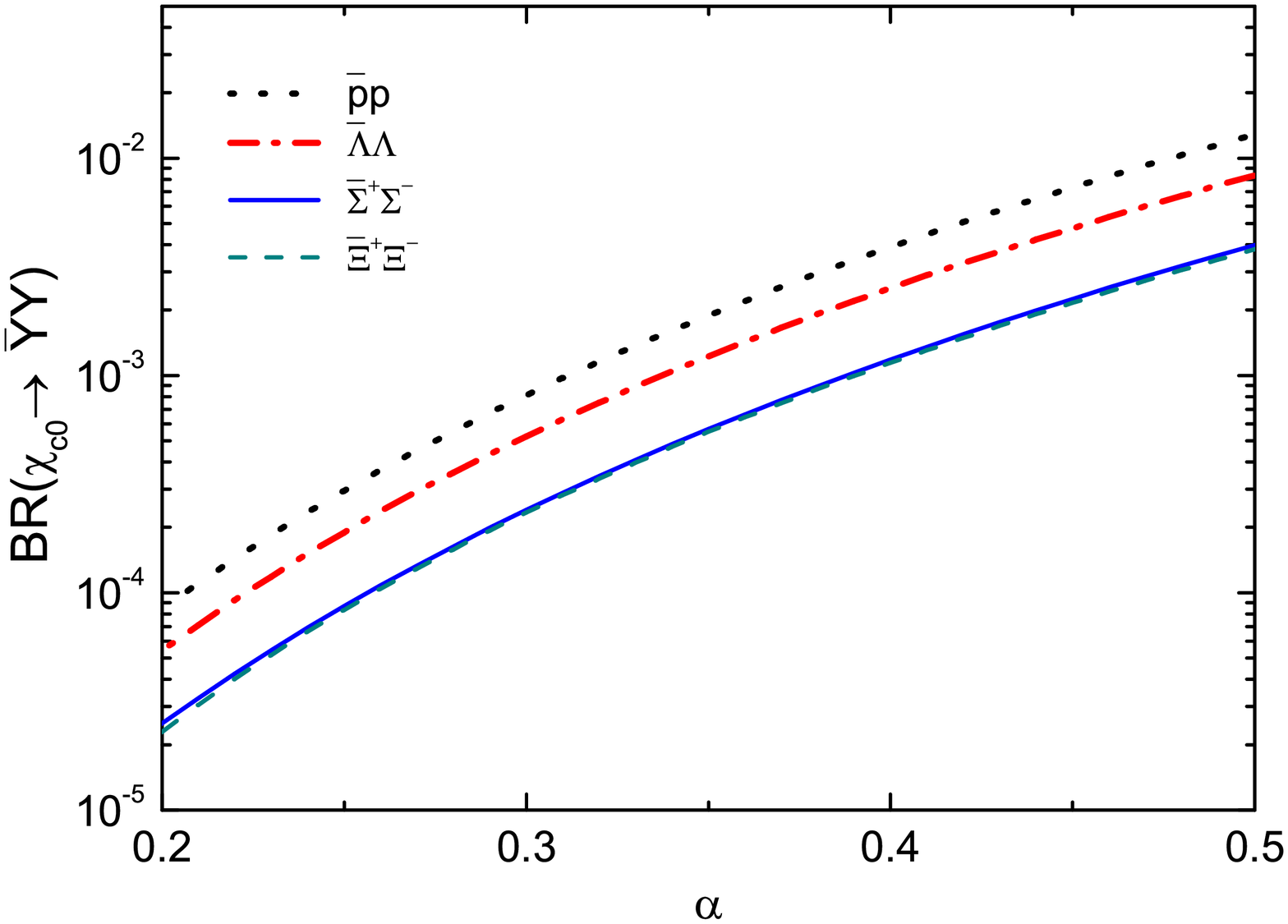}
\caption{ $\alpha$-dependence of the calculated branching ratios for
$\chi_{c0}\to Y\bar{Y}$. }\label{fig6}
\end{figure}

\begin{figure}[htb]
\includegraphics[width=0.5\hsize]{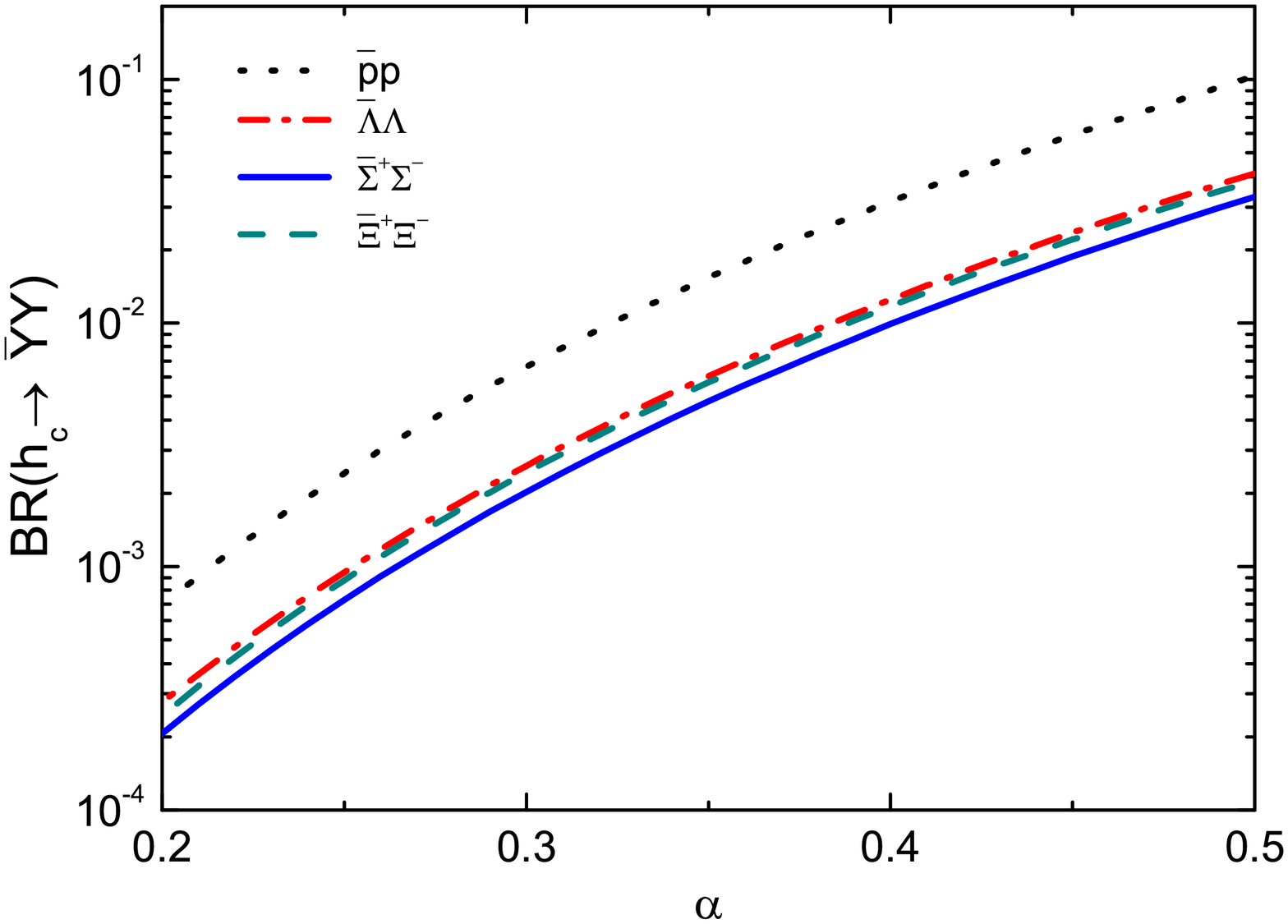}
\caption{ $\alpha$-dependence of the calculated branching ratios for
$h_c\to Y\bar{Y}$. }\label{fig7}
\end{figure}

In Figs.~\ref{fig5}, \ref{fig6} and \ref{fig7}, we also examine the
dependence of the results on the form factor parameter $\alpha$. The
adopted values of $\alpha$ are within a reasonable range and well
controlled by the available experimental data. Although some
uncertainties will be inevitably introduced by the phenomenological
form factor, the branching ratio fractions among the considered
channels turn out to be stable and less model-dependent. This
feature suggests that the branching ratio fractions are less
model-dependent in comparison with the absolute branching ratios. In
another word, although the model predictions for the absolute
branching ratios are lack of experimental constraints, thus, becomes
sensitive to the form factor parameter $\alpha$, we would expect
that the predicted branching ratio fractions among those considered
channels are less sensitive to it. As a consequence, any
experimental results for $\mbox{BR}(h_c\to Y\bar{Y})$ at ${\cal
O}(10^{-4})\sim {\cal O}(10^{-3})$ would imply the importance of the
hadron loop contributions to this helicity-selection-rule violation
transition.

In the $P$-wave charmonium decays the next higher Fock state
$c\bar{c} g$, i.e. the so-called color octet, will contribute at the
same order as the color singlet $c\bar{c}$ in the framework of
perturbative factorization method
\cite{Feldmann:2000hs,Cho:1995vh,Bolz:1997ez,Bodwin:1994jh,Wong:1999hc}.
This scenario may share the same intrinsic physics with those
intermediate charmed hadron loop transitions based on quark-hadron
duality argument, i.e. a manifestation of the same physics at either
quark-gluon level, or hadron level. Some qualitative discussions can
be found in Refs.~\cite{Cheng:2004ru,Isola:2001ar,Colangelo:2003sa}
and references therein. We address that since it is not easy to
handle these exclusive decays with the perturbative methods
considering the relatively low energy scale, the intermediate
charmed hadron loop transition would serve as a natural soft
mechanism in the numerical exploration.

\section{Summary}

In this work, with an effective Lagrangian method based on heavy
quark and flavor symmetry, we investigate the role played by the
intermediate charmed hadron loops in the processes $\eta_c,
\chi_{c0}, h_c\to Y\bar{Y}$, which are supposed to be highly
suppressed by the helicity selection rule. The results indicate that
the transitions via these kinds of loops as long-distance effects
can give significant contributions. This is a further test of the
mechanism for the evasion of helicity selection rule that we
proposed in Ref.~\cite{Liu:2010chi}. Although the model bares
uncertainties arising from the unknown coupling constants and form
factor parameter, the available data have provided a reasonable
control on the range of the form factor parameter values. Branching
ratios for some unmeasured channels can thus be predicted. Sizeable
data samples on charmonium decays accumulated at BEPCII and CLEO-c,
and future proton-antiproton annihilation data from $\mbox{P}$anda,
are expected to provide a great opportunity for revealing the
underlying mechanisms for charmonium helicity-selection-rule-evading
decays.

\section*{Acknowledgement}

This work is supported, in part, by the National Natural Science
Foundation of China (Grants No. 10675131 and 11035006), Chinese
Academy of Sciences (KJCX3-SYW-N2), and Ministry of Science and
Technology of China (2009CB825200).

\end{document}